\documentclass[graybox]{svmult}
\usepackage{graphicx} % Required for inserting images
\usepackage{multirow}
\usepackage{amsmath}
\usepackage{bm}
\usepackage{lineno}
\usepackage{amssymb}

\newcommand{\apj}{Astrophysical Journal}

\def\aap{A\&A}%
\def\aj{AJ}%
\def\araa{ARA\&A}%
\def\apj{ApJ}%
\def\apjl{ApJ}%
\def\mnras{MNRAS}%
\def\nat{Nature}%
\def\pasj{PASJ}%
\title*{Simultaneous Optical to X-ray Spectra of OJ 287: Insights into X-ray spectral changes and particle spectra}
\author{Pankaj Kushwaha\orcidID{0000-0001-6890-2236}}

\institute{Pankaj Kushwaha \at Department of Physical Sciences, Indian Institute of Science Education and Research Mohali, Knowledge City, Sector 81, SAS Nagar 140306, India \email{pankaj.kushwaha@iisermohali.ac.in}}

\begin{document}

\maketitle

\abstract{
OJ 287 is one of the most dynamic BL Lacertae objects that has exhibited behaviour representative
of the entire blazar class and is also one of the best sources with simultaneous multi-wavelength coordinated data. Motivated by
strong X-ray variability exhibited by the source, we systematically investigated the simultaneous optical to X-ray emission
 of the source with a focus on the spectral state of the lowest recorded X-ray flux state to understand the X-ray spectral
 changes. The optical-UV emission being synchrotron, the associated spectral variation is a direct reflection of the high-energy end of the
underlying particle spectrum and its power-law continuation to X-rays can drastically affect the X-ray spectrum without much change
in optical flux. Thus the combined optical to X-ray provides a potential tool to investigate and explore particle spectrum as well
as highest particle energies.  We report the finding of a power-law optical-UV spectrum with a photon spectral index of $2.71\pm0.03$ continuing to X-ray energies and accounting for this contribution at X-ray results in a photon spectral index of
$1.15-1.3$. We discuss the possible implications of this on X-ray spectral variations and the particle spectrum.
}

\section{Introduction}

Accretion-powered sources are the most dynamic astronomical sources in the sky exhibiting frequent and often
drastic changes in their observational behaviour, especially the compact systems. The observed behaviour and properties
of the latter imply a significant contribution from relativistic species, generally in the form of non-thermal components. Studies
exploiting different approaches to investigate these sources indicate quite broad similarities in some of the
observational properties (e.g. temporal: \cite{2015SciA....1E0686S,2016ApJ...822L..13K,2017ApJ...849..138K,2021Sci...373..789B,
2024ApJ...967L..18Z}, and references therein) while completely contrary in other properties (e.g. spectral: \cite{2017NatAs...1E.194P}
and references therein),
implying a very diverse range of extreme intrinsic conditions. 

Blazar is a sub-category within the active galactic nuclei (AGNs) referring to sources hosting a large-scale powerful
relativistic jet almost pointing in the direction of the Earth.
It comprised of traditionally labelled flat spectrum radio quasars (FSRQs) and BL Lacertae objects (BLLs). Their observed continuum radiation properties are drastically different 
from the typical galaxies and even the non-jetted AGNs, with an almost exclusively non-thermal component dominated
emission spread across the entire accessible electromagnetic (EM) windows from radio up to TeV gamma-rays (e.g. 
\cite{2010ApJ...716...30A,2017MNRAS.472..788G}).  The emission is characterised by a frequent and strong flux variability, strongly polarised 
at radio, optical, and even at X-ray in many sources (e.g. \cite{2022Natur.611..677L}) and it too varies in general with the source continuum variation (e.g.  \cite{2022PASJ...74.1041H,2023ApJ...957L..11G}). 

In terms of radiative output in different parts of EM 
windows, the broadband continuum exhibits a characteristic broad double-hump profile with a peak in near-infrared (NIR)
to X-ray energies and the other at sub-MeV to GeV energies (e.g. \cite{2022JApA...43...79K}). The low-energy component spreads from radio up
to X-rays and is widely considered to be the synchrotron emission from the relativistic electrons within the jet. The 
high-energy component, on the contrary, is still not well understood and is argued to be either due to relativistic leptons
(electrons/positrons) or hadrons (protons primarily). In the lepton-based scenario, the high-energy component results
from inverse Compton scattering of the surrounding photon fields while in the hadron-based explanation, it can result
from proton synchrotron or an outcome of cascade triggered due to proton-proton and/or proton-photon interactions.
The claim of association of neutrinos in a few blazars (e.g. \cite{2018Sci...361.1378I,2018Sci...361..147I,2022Sci...378..538I}) points towards hadronic contribution but current data support
minimal contribution via hadronic channels at MeV-GeV (e.g. \cite{2019NatAs...3...88G}), indicating
radio to GeV emission is primarily via leptonic channels.

OJ 287 is one of the most dynamic BL Lac objects located at the redshift of z=0.306. It is known for strong and frequent variability not only in flux/brightness but spectral and polarization as well (e.g. \cite{2018MNRAS.479.1672K,2019AJ....157...95G,2022MNRAS.513.3165K}). It exhibited
one of the most drastic spectral variations during 2016--2017 with spectral 
changes in all the observational windows of the electromagnetic spectrum,
especially at X-rays as reported in many previous works \cite{2017IAUS..324..168K,2021MNRAS.504.5575K,2018MNRAS.479.1672K,2022JApA...43...79K,2021ApJ...921...18K,2018MNRAS.480..407K,2022MNRAS.509.2696S,2021ApJ...920...12H}. Motivated by
the strong X-ray spectral variation as reported in many works, argued either
due to a new emission component (e.g. \cite{2018MNRAS.479.1672K,2021ApJ...921...18K,2021ApJ...920...12H}) or due
to the synchrotron component continuing to X-ray band (e.g. \cite{2022MNRAS.509.2696S}). In this proceeding, we report results
from the investigation of simultaneous optical to X-ray spectral variations with a focus on the lowest optical-UV and X-ray state of the source and X-ray spectral changes.

\section{Simultaneous Optical to X-ray Data}
\label{sec:data}
For transient and highly dynamic high-energy astronomical sources, the {\it Neils Gehrels Swift} observatory is the best facility
for broadband observations with autonomous
modes of operation capable of sampling a wide range of brightness states \cite{2004ApJ...611.1005G}. It hosts three astronomical payloads: the Ultra-Violet Optical Telescope (UVOT) with six astronomical filters covering optical to ultra-violet (UV) region, 
the X-ray Telescope (XRT) sensitive to X-ray photons of energies 0.2--10 keV, and the Burst Alert Telescope
(BAT) for photons of energies between 15 to 150 keV,  thereby offering simultaneous coverage from optical to hard X-ray bands of the electromagnetic spectrum. 
For our work here,  we have 
used the public archival data from two of the payloads: UVOT (Ultra-violet) and XRT (X-Ray Telescope) from the start of the
operation of the facility in 2005 to 2022 to study the spectral evolution of the BL Lac object OJ 287. 

Except for a few pointing observations, most of the XRT observations have a typical
exposure of around a kilo-second with an exposure of a few hundred seconds in a UVOT
filter. Such a typical exposure is sufficient for the spectral and temporal study of OJ 287 
most of the time by these instruments. The BAT is not so sensitive
 for temporal or spectral studies with this kind of exposure unless the source is too bright like a GRB and thus BAT is not considered (e.g. \cite{2020A&A...637A..55L}). 

The details of data UVOT and XRT reduction employed for this study are described
in \cite{2023arXiv230516144K}. In short, for flux density extraction from the UVOT
bands, a standard source region of $\rm 5''$ and an annular background region free
of any source of inner and outer radii $\rm 15''$ and $\rm 20''$ was employed.
For X-ray, a circular source region of $\rm 47''$ was used while the background was extracted using an annular region with inner
and other radii respectively of $\rm 65''$ and $\rm 90''$. The UVOT flux densities were then extracted using the task {\it uvotsource}
while XRT flux points were extracted by fitting the spectra within the XSPEC \cite{} with an
absorption-modified power-law (PL: $N(E) \equiv K (E/E_0)^{-\Gamma}$) and log-parabola model (LP: $N(E) \equiv K (E/E_0)^{-\alpha-\beta log(E/E0)}$). The best model out of the two was chosen based on the F-test value. For simultaneous optical-UV to X-ray 
spectral study, the UVOT files were converted to the respective PHA files using the HEASOFT task {\it uvot2pha} and the simultaneous fitting was performed in
the XSPEC. 

\section{Analysis and Results}
\label{sec:anRes}

\begin{figure}
    \centering
	\includegraphics[scale=0.67]{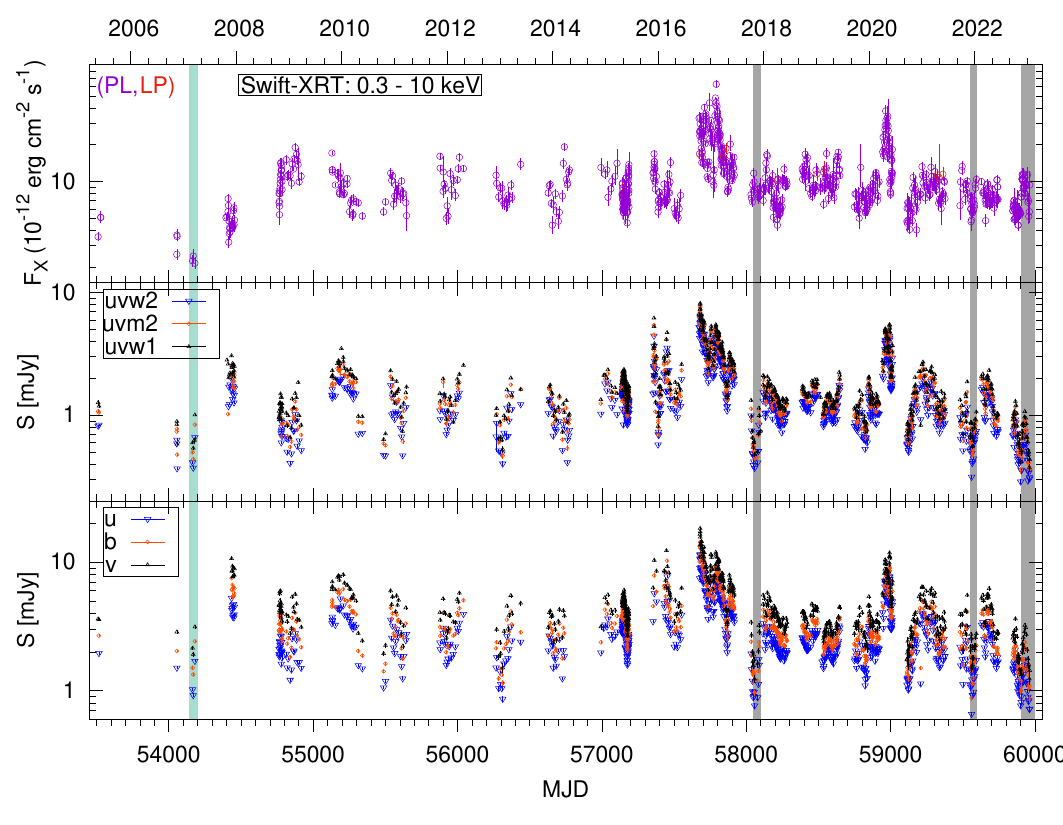}
    \caption{Simultaneous optical to X-ray light curve of OJ 287 from the {\it Swift}
    facility observation between 2005 to 2022 (ref \S\ref{sec:data}). The cyan shaded region is the focus of this work while the grey-shaded regions
    mark the periods with significant spectral changes in the broadband SED e.g. \cite{2017IAUS..324..168K,2018MNRAS.473.1145K,2018MNRAS.479.1672K,2021ApJ...921...18K,2022MNRAS.509.2696S}.}
 \label{fig:fig1}
\end{figure}

Figure \ref{fig:fig1} shows the simultaneous optical to X-ray light curve extracted using the pointed mode data from  the XRT and UVOT between 2005 and 2022 (MJD: 53510 -- 59970). In terms of variability, the XRT flux spans $>2$ orders of magnitude between the recorded minimum and maximum flux in the 0.3 -- 10 keV while UVOT bands during this period exhibit $>1$ order of magnitude in the flux density. The strong flux changes are associated with strong
spectral changes as reported in various studies (e.g. \cite{2017IAUS..324..168K,2018MNRAS.479.1672K,2021ApJ...921...18K,2022MNRAS.509.2696S}) and the 2016-2017 activity was also coincident
with the first ever reported very high energy (VHE; E $> 100$ GeV) activity of the source by the VERITAS observatory \cite{2017ICRC...35..650B}. 

In terms of simultaneous flux/brightness variability trend at optical-UV and X-rays, one can observe a peculiar feature -- a simultaneous low in both during MJD: 54160 -- 54180. Later though optical-UV has gone even lower but not the X-rays. 
Since many previous studies have reported synchrotron driving X-ray spectral changes (e.g. \cite{2022MNRAS.509.2696S,2020ApJ...890...47P,2001PASJ...53...79I} and references therein), we explored the optical-UV to X-ray spectral state of the lowest
optical-UV and X-ray state during this simultaneous low phase. For this,
we first searched the lowest X-ray flux and then the lowest optical-UV within
it, resulting in one (Obs-ID:00030901002). The details of criteria and fitting are given in \cite{2023arXiv230516144K} and the joint fitting was performed in
the XPSEC using the respective source, background, response, and auxiliary files. 

\begin{figure}
\centering
	\includegraphics[scale=0.205,angle=-90]{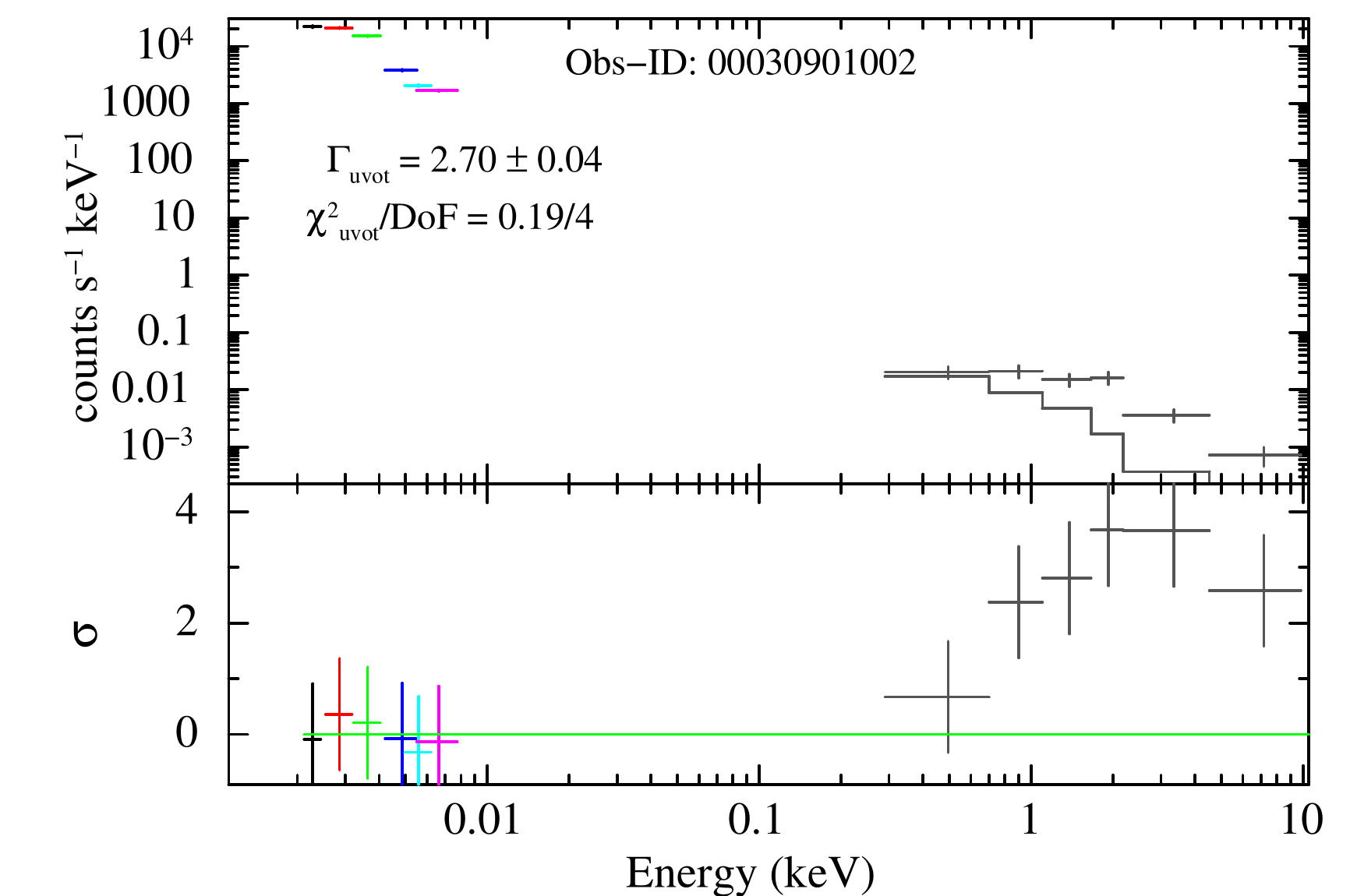}%\trim
	\includegraphics[scale=0.21,angle=-90]{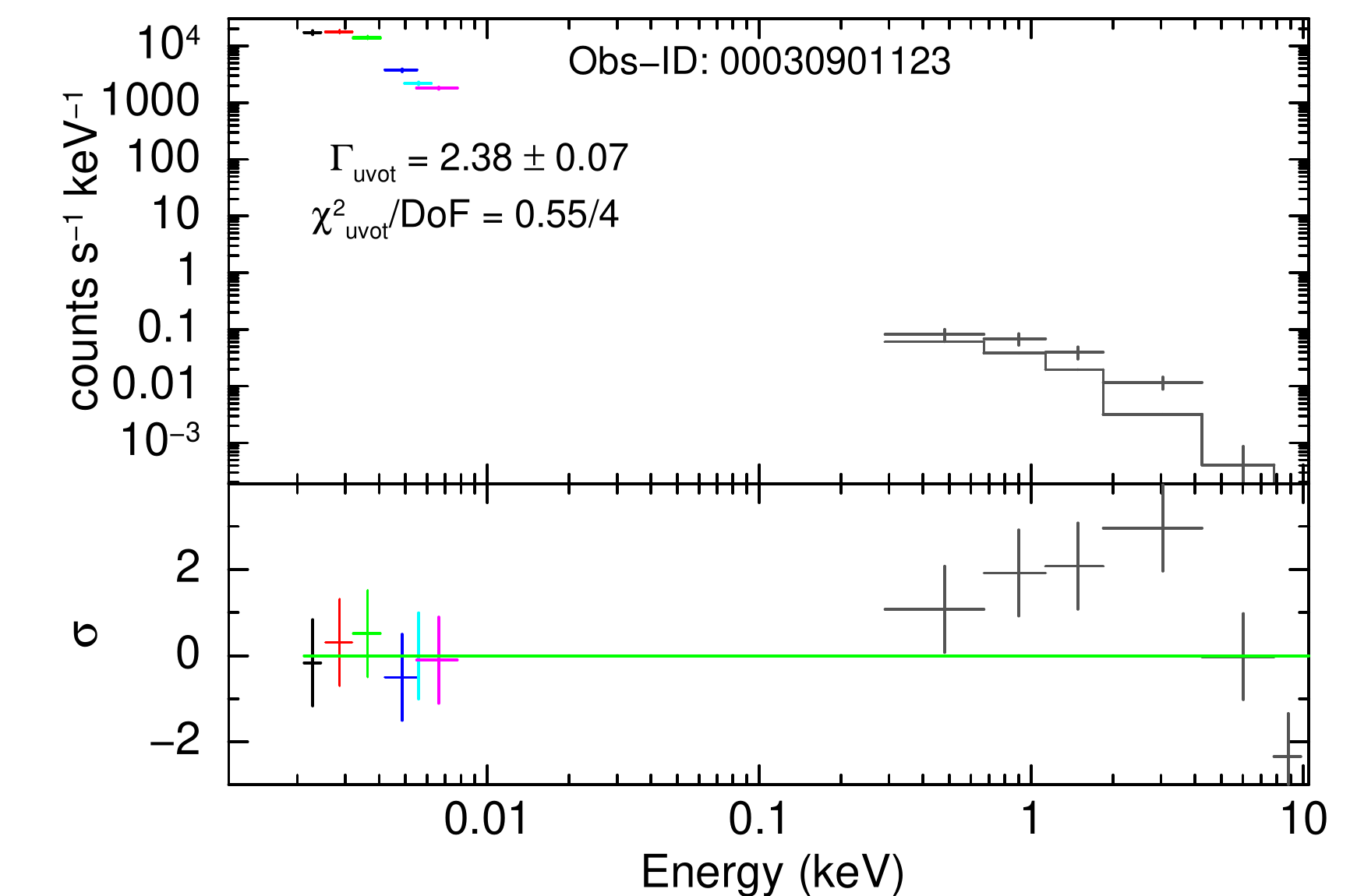}
	\caption{The best-fit power-law optical-UV spectrum and its comparison with the X-ray data for the lowest simultaneous optical-UV and X-ray phase: Obs-ID: 00030901002, and another similar optical-UV state with a different spectral index and also a higher X-ray flux: Obs-ID: 00030901123 (see \ref{sec:anRes}).}	
 \label{fig:fig2}
\end{figure}
 
First, we independently verified the optical-UV spectrum and found it consistent with a power-law
spectrum with a photon spectral index of $\rm \Gamma_{UVOT} = 2.70 \pm 0.04$  ($\rm \chi^2/DoF = 0.19/4$; DoF: Degree of Freedom). This best fit and its comparison with the X-ray data is shown in Figure \ref{fig:fig2} (Obs-ID:00030901002).  Similarly, we
independently verified the corresponding X-ray and found it consistent with 
a power-law with a photon spectral index of $\rm \Gamma_X = 1.41\pm0.15$ ($\rm
C_{stat} = 76.9/84$). Since the optical-UV best-fit contribution at X-ray is
lower/similar to X-ray data, assuming that optical-UV continues to extend to
X-ray, we performed a joint-fit and got an acceptable fit with a photon spectral
index of $\Gamma_{UVOT} = 2.71 \pm 0.03$ and $\Gamma_X = 1.22\pm0.20$. This is
the hardest reported X-ray spectrum of OJ 287 to the best of our knowledge and
consistent with the inferred X-ray spectrum at hard X-ray energies explored by
the Swift-BAT \cite{2020A&A...637A..55L}. 
The uncertainty is high due to low exposure leading to lower counts. 
However, joint modelling involving similar X-ray spectral states leads to
tighter constraints consistent with the current one (e.g. see \cite{2023arXiv230516144K} for more details). 

The above joint modelling and inferences were under the assumption that the power-law
optical-UV synchrotron spectrum extends unhindered to X-ray energies during this
simultaneous low flux state at optical-UV and X-rays. To firmly establish this,
we searched other optical-UV states having similar optical-UV flux and examined the X-ray spectra. Interestingly, we found an observation (ID:00030901123) having almost similar optical-UV flux in U-band but a very different spectrum, both at optical-UV and X-ray.
We found that this different X-ray spectrum can be simply reproduced with the 
corresponding synchrotron and the inferred low-hard X-ray spectrum inferred
for the low state as shown in Figure \ref{fig:fig3}, indicating an extended optical-UV spectrum
during very low optical-UV flux states.

\section{Discussion}\label{sec:Discuss}
X-ray emission from OJ 287 is generally attributed to the synchrotron self-Compton (SSC) i.e. inverse Compton
scattering of synchrotron photon (e.g. \cite{2001PASJ...53...79I,2013MNRAS.433.2380K}).
However, many studies have also reported significant contribution of the optical-UV synchrotron high-energy
tail to X-rays (e.g. \cite{2001PASJ...53...79I,2022MNRAS.509.2696S}).   Without any other contamination, the SSC spectral shape
is related to the spectral shape of the low-energy part of the particle distribution
and the low-energy cutoff (e.g. \cite{1996ApJ...463..555I,2006MNRAS.368L..52K}).

 As stated earlier, OJ 287 has shown significant X-ray spectral variability. With our focus on understanding X-ray spectral changes, especially  the
role of the optical-UV synchrotron component in driving these changes, we investigated the simultaneous optical-UV to X-ray variation of OJ 287 during the low optical-UV and X-ray state. Variability analysis of the light curve presented in Figure \ref{fig:fig1} shows that between 2005 and 2022 OJ 287 has exhibited drastic flux/brightness variability, by a factor of $>10$ at optical-UV and even more drastic at X-ray -- by a factor $>100$ between
the minimum and the maximum. The drastic flux variability is accompanied by an equally drastic spectral variability across the EM spectrum
as reported in previous multi-wavelength studies \cite{2017IAUS..324..168K,2021MNRAS.504.5575K,2018MNRAS.479.1672K,2022JApA...43...79K,2021ApJ...921...18K,2018MNRAS.480..407K,2022MNRAS.509.2696S,2021ApJ...920...12H,2021A&A...654A..38P}. Focussing on simultaneous\footnote{within the data and cadence} optical-UV and
X-ray variation, we found a duration: MJD 54160-54180 (cyan-coloured band in Figure
\ref{fig:fig1}) having the lowest simultaneous state in both and jointly investigated the optical-UV to X-ray spectrum. 

\begin{figure}
\centering
	\includegraphics[scale=0.8]{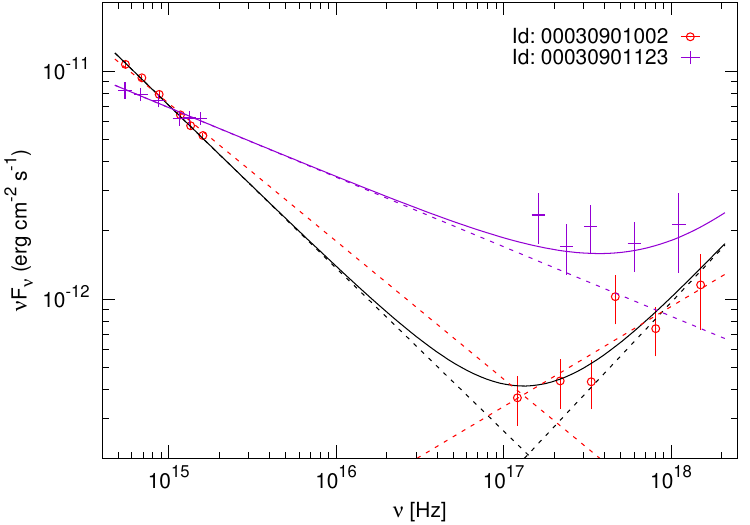}
	\caption{The simultaneous optical-UV to X-ray SED extracted from  for the lowest phase
	(Id: 00030901002) showing best-fit power-law shape from independent fitting
	(red dashed curves) and from the joint modelling (black dot-dashed curves).
	The purple data is the other optical-UV spectrum having almost the same UVW1-band flux along with the corresponding X-ray spectrum with curves showing independent optical-UV fit
	(purple dashed) and the reproduced of X-ray spectrum: purple-dashed optical-UV synchrotron + the inferred hardest X-ray spectrum during the lowest X-ray flux state  (see \S\ref{sec:anRes}).}	
 \label{fig:fig3}
\end{figure}

Jointly modelling the simultaneous optical-UV to X-ray spectrum of the lowest recorded X-ray flux state (see \S\ref{sec:anRes}), we found a power-law optical-UV spectrum with $\Gamma_{UVOT} = 2.71 \pm 0.03$ extending to X-ray energies (see Figure \ref{fig:fig2}-left).
Accounting for this contribution at X-ray led to a very hard X-ray spectrum
of $\Gamma_X = 1.22\pm0.20$.  This inference was factually verified and firmly established with our finding of another similar low-flux
optical-UV state with an altogether different
optical-UV and a much brighter X-ray flux as shown in Figure \ref{fig:fig2}-right. This different X-ray spectrum
is naturally reproduced as the sum of its optical-UV power-law synchrotron component and the lowest
inferred hard X-ray spectrum as shown in Figure \ref{fig:fig3}. 

Since the optical-UV continuum is synchrotron emission (e.g. \cite{2001PASJ...53...79I,2013MNRAS.433.2380K}) and directly
traces the high-energy part of the underlying particle distribution, the finding has two direct implications as far as the X-ray spectral changes are concerned in OJ 287:
(a) the optical-UV spectrum extends to X-ray energies and play significant role
in driving X-ray spectral changes during low-brightness phases and (b) the underlying particle spectrum governing the optical-UV synchrotron extends to much higher energies during low-brightness phases. 

Since the jet plasma is very rare (collision-less), the particle acceleration
and resulting spectrum are in general related to the size of confinement of the particle within the emission/acceleration region \cite{1984ARA&A..22..425H} and losses e.g. radiative, an extended particle
spectrum indicate a much larger emission region and also a relatively much lesser
radiative losses.  Thus the inferred low and hard X-ray spectrum could be a persistent low-lying component which may be the large-scale jet emission. The inferred optical-UV spectrum implies
a corresponding relativistic particle distribution with a power-law spectrum
of $\rm 4.4\pm0.6$. Similarly using the same relation, the inferred hard X-ray spectrum implies a particle spectrum of power-law index 1.15 -- 1.3. However, it should be noted
that the inferred X-ray particle spectrum may not be a true representative of
the underlying particle as such a hard state can also be reproduced via a particle
distribution with a power-law index closer to the one predicted in standard 
shock acceleration e.g., \cite{1996ApJ...463..555I,2006MNRAS.368L..52K,2023arXiv230516144K} with a relatively higher lower energy cutoff -- also consistent with
a larger acceleration/emission region.

\section{Summary and Conclusion}\label{sec:conSum}
OJ 287 has exhibited drastic spectral changes at X-rays which, in general, have been attributed to SSC but 
optical-UV synchrotron has also been reported to affect the X-ray spectrum in many previous studies. In the quest
to understand the role of the optical-UV synchrotron component in causing X-ray spectral changes, we systematically investigated the
simultaneous optical-UV and X-ray variability of the source with a focus on the simultaneous low-state in both. 

We found a simultaneous low optical-UV and X-ray phase and examined the lowest optical-UV to X-ray state.
First, we explored the optical-UV spectrum and found it consistent with a PL spectrum. We then accessed
this best-fit with the X-ray data by extrapolating the PL to  X-ray and found that the lowest energy X-ray data
is consistent with the best fit optical-UV spectrum. We then performed the joint fit of optical-UV to X-ray 
and inferred a very hard X-ray spectrum with a power-law photon index $\Gamma_X = 1.22\pm0.20$ -- the hardest reported
X-ray spectrum for OJ 287 to the best of our knowledge. We also found another observation with an almost
similar UVW1 band flux but a significantly different optical-UV and X-ray spectra with a much higher
X-ray flux. We showed that this X-ray spectrum can be naturally reproduced as the sum of its optics-UV
PL synchrotron component and the discovered hard X-ray spectrum, establishing
our assumption that optical-UV synchrotron spectrum during very low-optical-UV brightness extends to
much higher energies.

The inferred lowest flux yet hard X-ray state and the corresponding optical-UV synchrotron with a high-energy
tail imply a much larger acceleration region and could be a persistent emission component associated with 
the large-scale jet. The 
extended lowest-state optical-UV spectrum implies a particle distribution (high-energy part) with a power-law index of 
$\rm \sim 4.4$ while X-ray implies the low-energy part of a power-law index of $\rm \sim 1.15 -1.3$. It should be noted
that such a hard X-ray spectrum can be reproduced via the standard shock spectrum with a much higher low-energy cutoff in 
the particle spectrum and may not be reflective of an actual hard particle spectrum.

\section{Acknowledgement}
The author acknowledge support from the INSPIRE Faculty grant (DST/ INSPIRE/04/2020/002586) from the Department of Science  and Technology (DST), Government of India.

\end{document}